\documentclass[epjCONF]{svjour}
\usepackage{graphics}
\usepackage[varg]{txfonts} 
\session-title{Conference Title, to be filled}
\begin{document}
\title{On the lack of stellar bars in Coma dwarf galaxies}
\author{J. M\'endez-Abreu\inst{1,2}\fnmsep\thanks{\email{jairo@iac.es}} \and 
R. S\'anchez-Janssen\inst{3} \and J. A. L. Aguerri\inst{1,2} }
\institute{Instituto de Astrof\'isica de Canarias, Calle V\'ia L\'actea s/n,  E-38200 La Laguna, Tenerife, Spain 
\and Departamento de Astrof\'isica, Universidad de La Laguna, E-38205 La Laguna, Tenerife, Spain 
\and European Southern Observatory, Alonso de Cordova 3107, Vitacura, Santiago, Chile}
\abstract{We present a  study of the bar fraction  in the Coma cluster
  galaxies based on  a sample of $\sim$190 galaxies  selected from the
  SDSS-DR6 and observed with the Hubble Space Telescope (HST) Advanced
  Camera  for  Survey  (ACS).   The unprecedented  resolution  of  the
  HST-ACS images allows  us to explore the presence  of bars, detected
  by  visual classification, throughout  a luminosity  range of  9 mag
  ($ -23 \lesssim$ M$_{r} \lesssim  -14$), permitting us to study the
  poor known region of dwarf galaxies. We find that bars are hosted by
  galaxies  in a  tight  range of  both  luminosities ($ -22  \lesssim$
  M$_{r} \lesssim    -17$)    and   masses    ($10^9    \lesssim$
  M$_{*}$/M$_{\odot}$ $\lesssim 10^{11}$).   In addition, we find that
  the bar  fraction does  not vary significantly  when going  from the
  center to  the cluster outskirts, implying  that cluster environment
  plays a second-order role  in bar formation/evolution.  The shape of
  the bar  fraction distribution with  respect to both  luminosity and
  mass is well matched by the luminosity distribution of disk galaxies
  in  Coma, indicating  that bars  are  good tracers  of cold  stellar
  disks.}
%
\maketitle
\section{Introduction}
\label{intro}
Stellar bars are observed in optical images of roughly half of all the
nearby disk  galaxies \cite{barazza08,aguerri09}. This  fraction rises
to   about   59-62\%    when   near-infrared   images   are   analysed
\cite{marinovajogee07,menendezdelmestre07}.   It  is established  that
they appear naturally  in most simulations of galaxy  formation once a
dynamically cold and rotationally-supported disk is at place. However,
even if bars  are ubiquitous in the universe, it is  not clear yet why
one  galaxy  can exhibit  a  bar  structure  while another  apparently
similar does not. With  the recent advent of large galaxy surveys
statistical studies  of bar frequencies have  been possible.  However,
bar studies have  been usually restricted to luminous  galaxies due to
either the lack of spatial  resolution or because images were not deep
enough.   In  \cite{mendezabreu10}  we  attempt to  put  observational
constraints  on   the  internal  (mass)   and  external  (environment)
parameters   that  influence   bar   formation  by   carrying  out   a
comprehensive study of  the bar fraction in the  Coma cluster galaxies
throughout  a  wide  range   of  9  magnitudes,  covering  from  giant
ellipticals (M$_{r}  \sim -23$) to dwarf galaxies  (M$_{r} \sim -14$).
This research will also  provide us the luminosity/mass interval where
cold stellar disks are present in galaxies.

\section{Data and cluster membership selection}
\label{sec:1}
The  HST-ACS  Coma  Cluster  Treasury  Survey  \cite{carter08}  covers
$\sim$230  arcmin$^2$  with  21  ACS  pointings  ($\sim$3  $\times$  3
arcmin$^2$). At the distance of  the Coma cluster ($\sim$100 Mpc), the
resolution of  HST-ACS (0.1 arcsec) corresponds to  $\sim$50 pc.  This
gives  essentially  the   same  physical  resolution  as  ground-based
observations have in  Virgo and it will allow us  to resolve bars down
to  sizes of  $\sim$150 pc.   In addition,  the Coma  cluster  is also
covered by the SDSS, providing galaxy magnitudes in five bands ($u, g,
r, i, z$).

We  create our  catalog  of  sources using  the  SDSS-DR6.  The  steps
followed to obtain  our final sample of Coma  cluster members were: i)
download a catalogue of extended sources (m$_{r}$ $<$ 21, b/a $>$ 0.5)
from the SDSS-DR6 within a 5  arcmin radius from the position of every
ACS pointing. This  resulted in a sample of 477  galaxies, 104 of them
having  recession  velocities  available  from the  NED.   ii)  Select
galaxies with  velocities $\pm$3000  km/s (3$\sigma$) with  respect to
the Coma  redshift as cluster  members. All 104 galaxies  satisfy this
condition.   iii)  Visually  inspect  every galaxy  to  determine  its
possible   cluster   membership   based   on   its   morphology   (see
\cite{michardandreon08}).   From the  remaining  373 galaxies  without
redshift, 127 galaxies are cluster members following the morphological
criteria. iv) A color cut  condition was imposed.  Members should have
a $g-r$ color  less than 0.2 mag above the value  of the red sequence
fit.

Our final sample of Coma  secure members consists of 188 galaxies with
magnitudes in the range --23 $<$ M$_{r}$ $<$ --14.

\begin{figure*}[!ht]
\centering
\resizebox{0.99\textwidth}{!}{
  \includegraphics{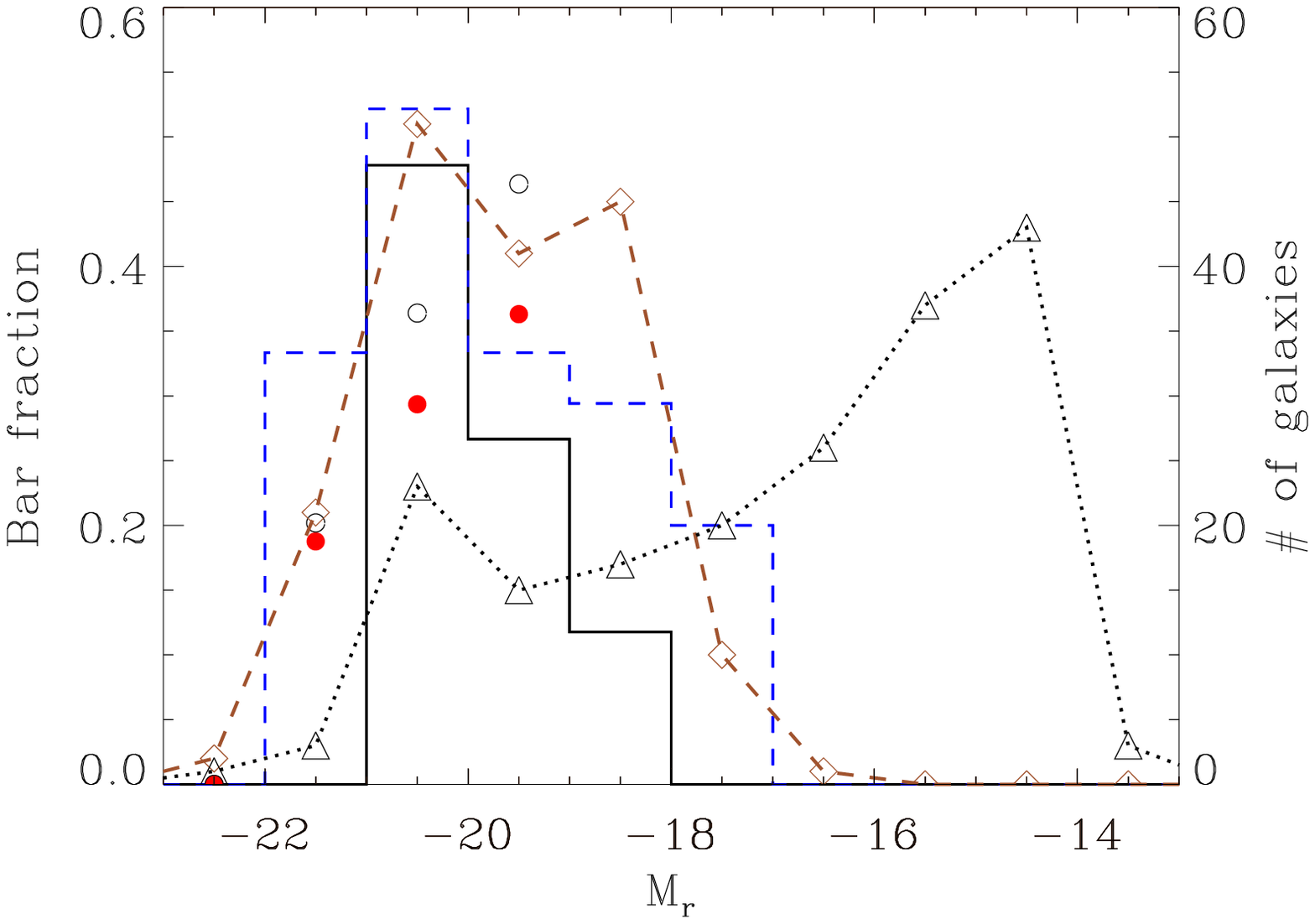}
\includegraphics{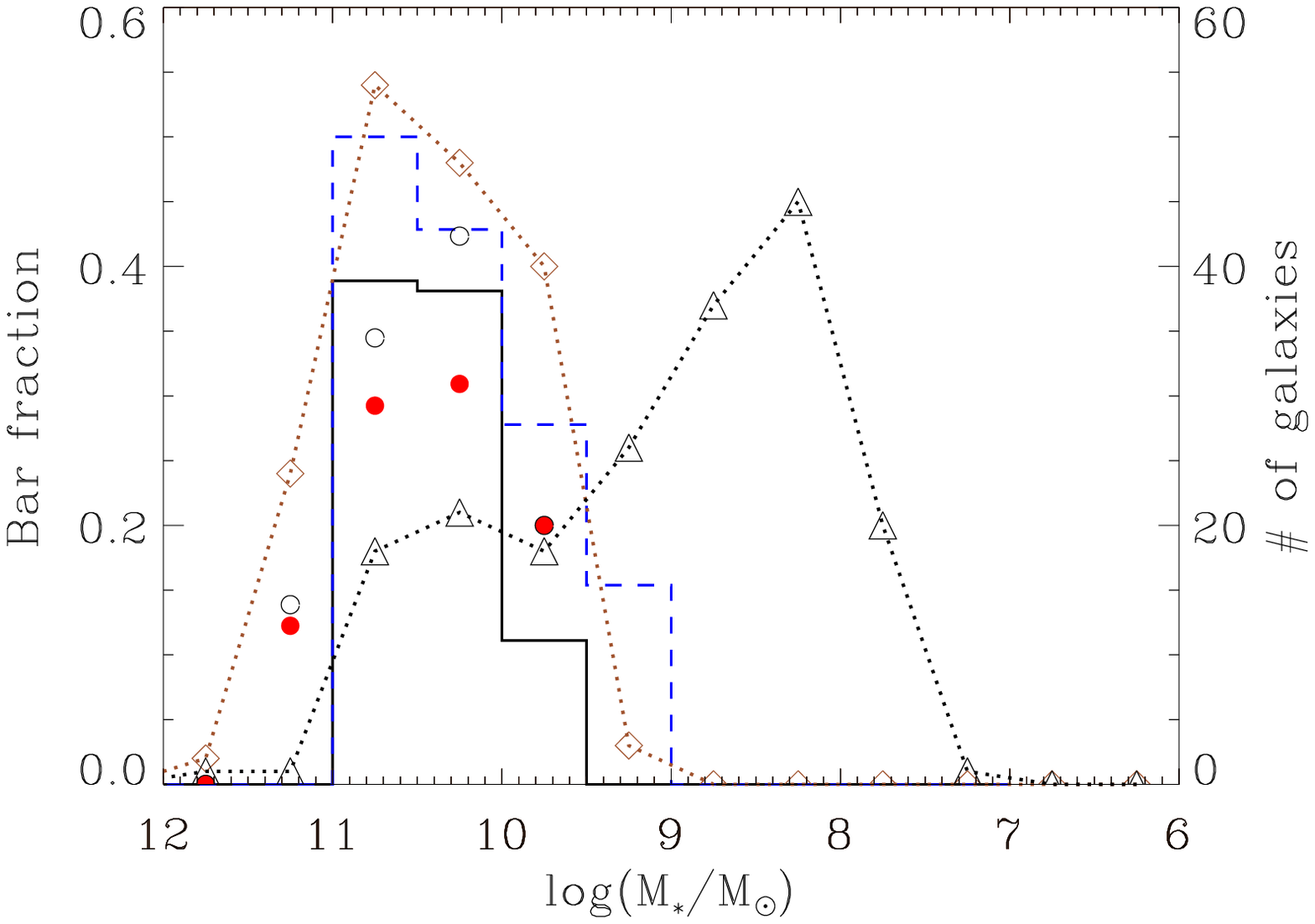}
}
\caption{Optical  bar  fraction  of  strong  (solid  black  line)  and
  weak+strong (dashed blue line) as  a function of the galaxy absolute
  magnitude  in   $r$  band  (left  panel)  and   galaxy  mass  (right
  panel).  Red  points and  black  circles  represent  the strong  and
  weak+strong bar  fraction using the  field sample of Aguerri  et al.
  (2009), respectively.  The number of galaxies per bin is represented
  with black  triangles. The disks luminosity  distribution by Michard
  \& Andreon (2008) is shown with brown diamonds.}
\label{fig:1}       
\end{figure*}

\section{Method and results}
\label{sec:2}
We visually classified all galaxies into strong barred, weakly barred,
and unbarred using the redder  available filter (F814W) of the HST-ACS
images.  Since our goal is to  understand where bars form, we have used
all galaxies, independently of their Hubble type, when calculating the
bar fraction.   Using this  definition, our bar  fraction turns  to be
$\sim$9\% and  $\sim$14\% depending if  only strong or also  weak bars
are included, respectively.

Fig. 1 shows the bar fraction as a function of the luminosity and mass
of the secure sample of galaxies in the Coma cluster. Independently of
the bar  strength, bars  are hosted  by galaxies in  a tight  range of
luminosities or masses.  There are  no strong bars in the Coma cluster
out of  the luminosity range between $-21  \lesssim$ M$_{r}$ $\lesssim
-18$.  These limits become $-22 \lesssim$ M$_{r}$ $\lesssim -17$ if we
also include weak bars. We find  the same behavior in the bar fraction
when  using the  galaxy mass:  bars exist  only in  a range  of masses
between  either 10$^{9.5}$  $\lesssim$  M$_{*}$/M$_{\odot}$ $\lesssim$
10$^{11}$   or  10$^{9}$  $\lesssim$   M$_{*}$/M$_{\odot}$  $\lesssim$
10$^{11}$  depending on whether  only strong  or strong+weak  bars are
considered, respectively.

\section{Discussion and Conclusions}
\label{sec:3}
If bars are tracers of cold  stellar disks, the presence of bars could
be useful to identify galaxies  with disks in clusters, and therefore,
the  distribution of  bar fraction  with the  galaxy  magnitude should
trace the shape of the disk galaxies' luminosity distribution. We have
computed the luminosity  distribution of morphologically selected disk
galaxies (from S0 to Sc) using the classification by Michard \& Andreon
(2008).  The  resulting luminosity distribution (Fig.  1) matches well
the  shape of  the bar  fraction distribution  in both  luminosity and
mass, confirming that bars are good tracers of disks.

Since our sample galaxies cover  both the center and infall regions of
the  Coma cluster,  we have  studied the  influence on  environment by
dividing  our   sample  into   internal  and  external   galaxies  and
calculating  the bar fraction  for every  subsample. We  have repeated
this procedure  for three values  of the separation distance  (0.5, 1,
and  1.5 Mpc)  from the  cluster center.   For the  smaller separation
distance, we found 14\% and 15\% of bars for the internal and external
subsamples, respectively.   The bar  fractions of both  subsamples are
14\%  and 17\%  when using  1 Mpc,  and 14\%  and 17\%  if we  use 1.5
Mpc.  Therefore, we  did  not  find differences  in  the bar  fraction
between the subsamples for  any separation distance, implying that the
cluster    environment   plays    a   second-order    role    in   bar
formation/evolution.

Few works have  investigated the dwarf realm in order  to look for the
presence of bars.   Lisker et al.  (2006), studying  a sample of dwarf
galaxies in the  Virgo cluster, claim that dwarf  ellipticals with and
without disks represent  two distinct types of galaxies,  and show how
the fraction of dwarfs  with disks decreases dramatically for galaxies
below M$_{B}$ $\sim$ --16.  Therefore, even if the non-presence of bars
in our low-luminosity galaxies could be due to the heating of the disk
component or due to its absence,  our result support that of Lisker et
al. (2006),  and we suggest  that no disks  are present in  Coma dwarf
galaxies below M$_{r}$ $\sim$ --17.   This result is also supported by
the recent work  of S\'anchez-Janssen et al. (2010)  where they found,
for  a large  sample  of galaxies  in  the field,  that disks  becomes
thicker for galaxies with M$_{*}$/M$_{\odot}$ $\leq$ 2$\times$10$^9$.

\end{document}